\documentclass[useAMS,usenatbib,onecolumn
]{mn2e}
\usepackage{graphicx}
\usepackage[section]{placeins}

\usepackage{times} 
\usepackage{amsmath}
\usepackage{amssymb}

\begin{document}

\title[MOND as modified inertia]{Towards an interpretation of MOND as a modification of inertia}
\author[Fathi Namouni]{Fathi Namouni\thanks{E-mail:
namouni@obs-nice.fr} \\
Universit\'e de Nice, CNRS, Observatoire de la C\^ote d'Azur, CS 34229, 06304 Nice, France}

\date{Accepted 2015 June 8.  Received 2015 June 5; in original form 2015 March 31}

\maketitle

\begin{abstract}
We explore the possibility that Milgrom's Modified Newtonian Dynamics (MOND) is a manifestation of the modification of inertia at small accelerations. Consistent with the Tully-Fisher relation, dynamics in the small acceleration domain may originate from a quartic (cubic) velocity-dependence of energy (momentum) whereas gravitational potentials remain linear with respect to mass. The natural framework for this interpretation is Finsler geometry.  The simplest static isotropic Finsler metric of a gravitating mass that incorporates the Tully-Fisher relation at small acceleration is associated with a spacetime interval that is either a homogeneous quartic root of polynomials of local displacements or a simple root of a rational fraction thereof. We determine the weak field gravitational equation and find that Finsler spacetimes that produce a Tully-Fisher relation  require that the  gravitational potential be modified. For an isolated mass, Newton's potential $Mr^{-1}$ is replaced by $Ma_0\log (r/r_0)$ where $a_0$ is MOND's acceleration scale and $r_0$ is a yet undetermined distance scale. Orbital energy is linear with respect to mass but  angular momentum is proportional to $ M^{3/4}$.  Asymptotic light deflection resulting from time curvature is similar to that of a singular isothermal sphere implying that space curvature must be the main source of deflection in static Finsler spacetimes possibly  through the  presence of the distance scale $r_0$ that appears in the asymptotic form of the gravitational potential. The quartic nature of the Finsler metric hints at the existence of an underlying area-metric that describes the effective structure of spacetime.
 \end{abstract}


\section{Introduction}
Milgrom's theory of Modified Newtonian Dynamics (MOND) was put forward to  model phenomenologically the velocity excess of galactic rotation curves \citep{Milgrom83}. In contrast to the dark matter theory, MOND assumes that only luminous galactic matter is responsible for the observed velocities but that Newtonian dynamics breaks down where stellar motion involves accelerations smaller than a constant, $a_0$, that is of order $cH$ where $c$ is the light speed and $H$ the Hubble constant. Newton's second law then assumes the modified quadratic form $a^2a_0^{-1}\propto F_g^N$ where $a$  and $F_g^N$ are the inertial and Newton's gravitational accelerations respectively. As a phenomenological theory, MOND's ability to fit galaxy rotation curves is well established \citep{Mcgaugh08,Milgrom12}. This success is also paralleled by the theory's inability to model the dynamics of galaxy clusters without requiring an unseen matter additive \citep{Famaey12}, and explain the morphology of galaxy cluster encounters \citep{Clowe06,Bradac08} that seem to strengthen the case for dark matter. These failings are not necessarily intrinsic to MOND's basic phenomenology as much as they are to its available formulations that may not describe the physical theory from which the scaling relations, motivated mainly by flat galactic rotation curves, originate.

In  search for the physical foundation of MOND, the modification of Newton's second law has been viewed as an indication of a modification of gravity as well as an indication of a modification of inertia. Modifying gravity entails searching for theories where motion is described by $a=F_g$ and $F_g$ is no longer the Newtonian gravitational acceleration but is related to it by the scaling relation $F_g=(a_0F_g^N)^{1/2}$ for small accelerations. The first such example is the widely used low-energy lagrangian-based theory of \cite{Bekenstein84} that modifies the gravitational Poisson equation nonlinearly. Various covariant theories for the modification of gravity along these lines have been formulated subsequently and allowed the application of the MOND paradigm to gravitational lensing and large scale structure formation.  Modifying inertia amounts to explaining how for small accelerations, the inertial resistance of a body to an external force becomes $a^2a_0^{-1}$ instead of the usual Newtonian linear scale-free behavior.  Gravity in the weak field limit is treated just like any other force and given by the standard Poisson equation. Attempts at deriving a formulation for the modification of inertia have not been successful because  deriving a lagrangian based theory satisfying Galilean invariance requires the theory to be non-local \citep{Milgrom94}. Together with the nonlinearity of MOND, this property  makes finding a working formulation as a modification of inertia more difficult.   More recently, it has been suggested that MOND phenomenology may be interpreted as the manifestation of spacetime scale invariance for small accelerations \citep{Milgrom09,Milgrom14}. 

In this work, we seek a formulation of MOND as a modification of inertia in a different way. We base our search exclusively on the observations that MOND was initially conceived for and explains best, namely, flat galaxy rotation curves. We assume that Milgrom's original  phenomenological  scaling  $a^2a_0^{-1}\propto F_g^N$ and the implied  quadratic inertial resistance do not necessarily reflect a new universal dependence on the acceleration $a$ in general but only the modification of inertia for nearly-circular orbits from $v^2r^{-1}$ at large accelerations to $v^4r^{-2}a_0^{-1}$ at small accelerations where $r$ and $v$ are the circular radius and velocity respectively. Although this inertial scaling when equated with Newton's gravitational acceleration $GMr^{-2}$ gives the flat rotation curves as originally sought by Milgrom, we show in this work that a metric description of this inertial change requires the gravitational acceleration to be changed to $GMa_0r^{-1}$ where $G$ is the gravitational constant. Although our starting point is the original MOND scaling, the approach we lay down in this work is based on metric spacetimes and at no point do we refer to the notion of force. The phenomenological scaling serves only as a guide for the kind of dynamical behavior we seek.

An inertial scaling proportional to the fourth power of velocity is indicative of a possible Finsler structure of spacetime. In effect, in Finsler geometry, the  proper time interval is no longer  the square root of a quadratic polynomial of  local displacements as in Riemann geometry. Instead, the spacetime interval may be a general homogeneous function of order 1 with respect to  local displacements that can be chosen to produce a quartic dependence. In this context, 
our scope of implementing the Tully-Fisher relation for nearly circular orbits around a spherically symmetric mass distribution translates into searching for the simplest static isotropic Finsler metric that produces an inertia proportional to the fourth power of velocity for circular gravitational motion.

What we seek to do for MOND  in a Finsler spacetime is analogous to the following programme: knowing Kepler's third law, one asks whether it is possible to write down the Riemann metric of a static isotropic  mass distribution where inertia is proportional to the second power of velocity.  If the mass distribution is concentrated in a single point, the sought-for result is the Schwarzschild metric. It is elementary, however, that Kepler's law does not define completely the Schwarzschild metric. Writing the Riemann spacetime interval as $d\tau^2=b(r) dt^2-a(r)dr^2-r^2d\phi^2$ in the standard spherical coordinates, one can derive the time potential $b(r)$ from Newtonian theory but not  the space potential, $a(r)$  \citep{Gruber88}. The space potential may be determined from the behaviour of light deflection as both space and time curvatures contribute to the effect (equally). Incidentally, we remark that the time potential $b(r)$ obtained from Newtonian theory is exactly that which enters the Schwarzschild metric regardless of the assumed weak field approximation of Newtonian dynamics. 

Transposing this programme in the context of the MOND quartic scaling in a Finsler spacetime is more complex for two reasons. The first is the presence of more than two potentials in the metric's expression (or more precisely in the Finsler function as explained in the next section). The second reason is the fact that  light deflection is poorly known  in the context of MOND as was mentioned above regarding gravitational lensing in merging galaxy clusters. We will however be able to determine the Finsler time potential that allows us to characterize  MOND dynamics in the weak field approximation for massive motion. By considering light deflection in the same weak field approximation, we will be able to determine that asymptotic deflection by a static isotropic mass distribution is caused mainly by the space potential unlike that of general relativity. 

The paper is organized as follows: in section 2, we briefly review the physical applications of  Finsler geometry and recall its main features  that will be used in this work. In section 3, we derive the properties of Finsler spacetime metrics that produce MOND-like inertia, write down the corresponding  weak field gravitational equation based on the Tully-Fisher relation at small acceleration and study its properties. Light deflection is analyzed in the weak field approximation  in  Section 4. We discuss our findings and their possible implications in Section 5.  In the following, the gravitational constant and the light speed are set to unity.

\section{Modification of inertia and Finsler spacetimes}
In classical physics, the non-relativistic linear dependence of  momentum on velocity and the quadratic dependence of kinetic energy on velocity stem from the Riemannian structure of spacetime for which the proper time interval is given by the root of a quadratic form of local displacements as $d\tau=(g_{\mu\nu}\ dx^\mu dx^\nu)^{1/2}$ where  $g_{\mu\nu}$ is the  metric tensor. That Nature has chosen this structure is obviously well established within the experimental range that current physics can reach. Nonetheless, different spacetime structures have been envisaged in various contexts: \cite{Riemann} considered the possibility that  the spatial distance interval could be  better described as a quartic polynomial of the local displacements rather than the standard quadratic form. The hypothesis was investigated extensively with proper time intervals of the form $d\tau=(h_{\kappa\lambda\mu\nu}  dx^\kappa dx^\lambda dx^\mu dx^\nu )^{1/4}$, where $h_{\kappa\lambda\mu\nu}$ is the gravitational 4-tensor, in the context of the metric  tests of General Relativity \citep{Roxy79,Roxy91,Roxy92a,Roxy92b}. More recently, a non-Riemannian structure of spacetime appeared in an attempt to circumvent local Lorentz and CPT invariance by assuming that the local laws of physics are not invariant with respect to the full Lorentz group of transformations \citep{Cohen06}. Such limited invariance describes  anisotropic spacetimes with the proper time interval $d\tau = (\eta_{\mu\nu}dx^\alpha dx^\nu)^{(1-b)/2}(n_\rho dx^\rho)^b$ where  $|b|\ll 1$, $\eta_{\mu\nu}$ is the Lorentz metric and the vector $n_\rho$ is position independent  \citep{Bogos99,Gibbons07}.  String fluid motion in area-metric spacetimes that models localized massive particles may also be described by a non-Riemannian quartic metric \citep{Punzi09a}.  Lastly, exotic Finsler spacetimes have been considered as a possible premise for gravitation theories but their physical foundations are still lacking \citep{Chang08,Li10}.

The relevant framework for describing such non-Riemannian modifications of inertia is Finsler geometry \citep{BaoChernShen} wherein the spacetime interval is written as $d\tau=F(x , dx) $ and $F$ is a positive homogeneous function of degree 1 with respect to  local displacements $dx $ in order to ensure proper time reparametrization invariance. A spacetime metric $f_{\mu\nu}$ may be derived from the Finsler length function $F$ as:
\begin{equation}
f_{\mu\nu}(x,v)=\frac{\partial^2F^2}{2\partial v^\mu \partial v^\nu }, \label{finslermetric}\   \mbox{and} \ 
f_{\mu\nu}(x,v) v^\mu v^\nu = F^2(x,v),
\end{equation}
where the latter property stems from the 0-homogeneity of $f_{\mu\nu}$ with respect to $v$. As the expressions of $f_{\mu\nu}$ are bound to be cumbersome because of the velocity dependence, it is common to use only  the Finsler length function $F$. Geodesics of the Finsler spacetime are obtained from the usual length action:
\begin{equation}
S=\int F\left(x, \frac{dx }{d\lambda}\right)d\lambda=\int \left[f_{\mu\nu}\left(x, \frac{dx }{d\lambda}\right)\frac{dx^\mu }{d\lambda}\frac{dx^\nu }{d\lambda}\right]^\frac{1}{2} d\lambda. \label{actionS}
\end{equation}
The existence of a well-defined  causal structure in the Finsler spacetime requires the  Finsler function to be of the form $F=L^{1/n}$ where $L$ is a positive homogeneous function of degree $n\geq 2$ \citep{Pfeifer11,Pfeifer14}.  

We approach the problem of the interpretation of MOND as a modification of inertia by considering a static spherical system in the small acceleration domain, and seeking the simplest static isotropic Finsler metrics that produce the Tully-Fisher relation, $v^4\propto M(r)$, for nearly circular orbits. It is clear from the outset that this condition will not define the Finsler metric entirely. The equations of motion we derive  in the weak field approximation will be sufficient to ascertain the similarities and differences between the gravity and inertia modification interpretations of MOND in the small acceleration domain. In Section 4, we proceed further and examine the properties of asymptotic light deflection.   

On account of the causality requirement that the Finsler function must be of the form $F=L^{1/n}$ where $L$ is $n$-homogeneous  with $n\geq 2$, $L$ must be a polynomial  or a rational fraction  of degree $n$ with respect to local displacements. Limiting the search to isotropic spacetimes restricts $n$ to even numbers. The simplest two possibilities are $n=2$ and  $n=4$. For $n=2$, $L$ must be a rational fraction, the simplest being the ratio of two polynomials respectively of degree 4 and 2. Riemann geometry is obtained if $L$ is a second degree polynomial. For $n=4$, the simplest $L$ is a polynomial of degree 4. On account of spherical symmetry, $F$ is function of only the spherical radius $r$, $dt$, $dr$, and the solid angle element $d\Omega=(d\theta^2+\sin^2\theta d\phi^2)^{1/2}$  \citep{Mccarthy93,Pfeifer12}. Furthermore, in a static spherically symmetric spacetime, energy and angular momentum are constants of motion. This stems from of the Lagrange equations of the proper length action ($\ref{actionS}$): $d(\partial_{v_\mu} F)/dt-\partial_{x^\mu} F=0$. For isntance,  as $\partial_t F=0$ then $\partial_{v_t}F=E$ is constant.

\section{Static isotropic Finsler spacetimes for MOND}
\subsection{Quartic metrics}
Quartic Finsler functions that produce  MOND-like inertia must not have time and space cross terms. To see this, consider the gravitation-free $F^4= v_t^4-  \alpha_{tx} v_t ^2 v^2- \alpha_{x} v^4$ where $\alpha_{tx}$ and $\alpha_{x}$ are constants. The non-relativistic kinetic energy and linear momentum are expanded to fourth order in velocity as: $E_k=  \partial_{v_t}F-1=\alpha_{tx} v^2/4 + ({9 \alpha_{tx}^2}/{32} + {3\alpha_{x} }/{4}) v^4$ and $p_i=-\partial_{v_i}F=\alpha_{tx} v_i/{2} +({3 \alpha_{tx}^2}/{8} + \alpha_{x}  ) v^2 v_i$ where $i$ is a space index and $v_i=dx^i/dt$. A quartic energy dependence requires $\alpha_{tx}=0$. Similarly, the most general static isotropic  metric that modifies the small-velocity energy law from $v^2$ to $v^4$  is given by:
\begin{equation}
F_q^4=\left(\frac{{\rm d}\tau}{{\rm d}\lambda}\right)^4=   \beta v_t^4 - (\gamma-\alpha^2) v_r^4 - ( \alpha v_r^2 + r^2 v_\Omega^2)^2\label{quarticmetric},
\end{equation}
where $v_\mu=dx^\mu/d\lambda$ and the potentials  $\alpha$, $\beta$ and $\gamma$ are functions of the radial variable $r$ only.  The coefficient of $r^4v_\Omega^4$ was set to unity by choosing spherical coordinates that describe space as nested spheres of radius $r$ for constant time $t$ and radius $r$. We note that the original  MOND scaling  applies only  to angular motion so that requiring the time and radius cross term to vanish is motivated by our view that inertia isotropy  should take the same form in all space directions. 

To proceed further, we first consider the spacetime defined by $F_q$ with $\gamma=\alpha^2$ as the associated effective gravitational potential may be written explicitly. Hence: 
\begin{equation}
F_q^4=\left(\beta^\frac{1}{2}v_t^2-\alpha v_r^2-r^2v_\Omega^2\right)\left(\beta^\frac{1}{2}v_t^2+\alpha v_r^2+r^2v_\Omega^2\right)\label{quarticmetric2}.
\end{equation}
In order to apply the Tully-Fisher relation to circular orbits, we determine the timelike and null geodesic equations directly from the energy  and angular momentum constants.  Owing to spherical symmetry, orbital motion lies in a constant plane that we define as $\theta=\pi/2$. As a consequence, the solid angle $\Omega$ is replaced  by the longitude $\phi$.  Choosing the affine parameter $\lambda =\tau$ for massive particles, the constants of motion are given by  $E= \partial_{v_t}F_q=\beta v_t^3$ and $J=-\partial_{v_\phi}F_q= r^2 v_\phi ( \alpha v_r^2 + r^2 v_\phi^2)$. Substituting the constants into the definition of proper time (\ref{quarticmetric}) yields the orbital equations in terms of  effective potentials as follows:
 \begin{eqnarray} 
\alpha v_r^2&=& (\beta^{-\frac{1}{3}} E^\frac{4}{3}-1)^\frac{1}{2}-\frac{J^2}{r^2(\beta^{-\frac{1}{3}} E^\frac{4}{3}-1)}\equiv -V_{\rm eff,m} \ \ \ \mbox{for massive particles},\\
\alpha v_r^2&=& \beta^{-\frac{1}{6}} E^\frac{2}{3}-\frac{J^2}{r^2\beta^{-\frac{1}{3}} E^\frac{4}{3}}\equiv -V_{\rm eff,0} \ \ \ \mbox{for photons}.
\end{eqnarray}
Circular orbits for massive particles are the energy minima of the effective potential and satisfy:  $V_{\rm eff,m}(r_{\rm circ})=0=\partial_rV_{\rm eff,m}(r_{\rm circ})$. Combining the two equations and using the expression of the circular velocity $v_{\rm circ}(r)=rv_\phi$, we get:
\begin{equation}  
\partial_r \log \beta(r)=\frac{4 v_{\rm circ}(r)^4}{r\left[1+v_{\rm circ}(r)^4\right]} \label{potquartic}.
\end{equation}
For comparison, the time potential, $b(r) $, of a static spherically symmetric system with a Riemann metric $d\tau^2=b dt^2-adr^2-r^2d\phi^2$ satisfies the  equation $\partial_r \log b=2v_{\rm circ}^2/{r\left(1+v_{\rm circ}^2\right)}$.  We may determine a simpler equation for the potential $\beta$  by noting that the Tully-Fisher relation is verified only at low energy ($v_{\rm circ}\ll 1$). Assuming further that spacetime is flat in the absence of matter,  we may use in the weak field  limit the fact that $|\beta -1|\ll 1$. Equation (\ref{potquartic}) then becomes $\partial_r\beta= 4 M(r)a_0r^{-1}$ where the Tully-Fisher relation, $v_{\rm circ}^4=M(r)a_0$ was used to eliminate the circular velocity.  In terms of the matter density $\rho=(4\pi r^2)^{-1}\partial_rM$, the weak field gravitational equation reads: 
\begin{equation}
r^{-2}\partial_r\left(\frac{r \partial_r\beta}{2a_0}\right)=8\pi\rho(r). \label{gravquartic}
\end{equation}
Following the same approximations but using the Kepler velocity relation instead of the Tully-Fisher relation, the time potential $b(r)$ of the Riemann metric satisfies the  Poisson equation $ r^{-2}\partial_r\left(r^2 \partial_rb\right)=8\pi  \rho(r)$ in the Newtonian limit. To determine an expression of the time potential $\beta(r)$, regardless of the weak field approximation, requires the knowledge of the exact 
velocity relation or the full gravitational equation that replaces the Einstein equation in the small acceleration limit. Lacking both elements, we will continue our study of the Finsler metric using equation (\ref{gravquartic}). The weak field gravitational equation should be able to yield the asymptotic behavior in the small acceleration limit (i.e. the dynamics far away from an isolated mass) in the same way the solution of the Poisson equation for the Riemann metric encapsulates the fundamental features of the dynamics of an isolated mass and incidentally yields the exact expression the time potential $b(r)$ regardless of the weak field approximation.

For a point mass, equation (\ref{gravquartic}) yields $\beta(r)= 1+4M a_0\log (r/r_0)$ where $r_0$ is a constant that may depend on $M$ and $a_0$. Whereas the constant $r_0$  does not influence qualitatively the properties of massive motion, its possible dependence on $M$ and $a_0$ influences light deflection as we explain in Section 3.  The energy and angular momentum of circular orbits are given as:
\begin{eqnarray} 
E_{\rm circ}&=& \frac{1+ 4 Ma_0 \log (r_{\rm circ}/r_0)}{[1- Ma_0 - 4 Ma_0 \log (r_{\rm circ}/r_0)]^{3/4}} 
\sim 1  + Ma_0 [3/4+ \log (r_{\rm circ}/r_0)], \ \ \ \ \ \mbox{for }Ma_0\ll 1, \nonumber\\
J_{\rm circ}&=&\frac{ (Ma_0)^\frac{3}{4}r_{\rm circ}}{(1- Ma_0 - 4 Ma_0 \log (r_{\rm circ}/r_0)]^{3/4}}\sim
 (Ma_0)^\frac{3}{4}r_{\rm circ}\left(1  + 3 Ma_0 [1/4+ \log (r_{\rm circ}/r_0)]\right),\ \ \ \ \  \mbox{for }Ma_0\ll 1. \nonumber
\end{eqnarray}

A first interesting result here is that modifying inertia through a non-Riemannian structure of spacetime requires a modification of gravity even though gravity remains linear with respect to mass at low energy. For comparison, the weak field modified gravity MOND potential is of the form $(Ma_0)^{1/2}\log(r/r_0)$,  orbital energy is $\propto (Ma_0)^{1/2}$ and angular momentum is $\propto (Ma_0)^{1/4}r_{\rm circ} $.

 The timelike effective potential, $V_{\rm eff,m}$, of a point mass is shown in Figure (1,left). Motion is always bound regardless of the unknown potential $\alpha$ and the sign of orbital energy $E-1$, as the condition $\beta^{-\frac{1}{3}} E^\frac{4}{3}-1> 0$  implies $r<r_0\exp[(E^4-1)/Ma_0]$. The potential's details at small radii and in particular its unstable circular orbit that resembles that of the Schwarzschild timelike effective potential are unimportant at this stage as the small acceleration approximation is no longer valid and the metric structure should tend to that of a Schwarzschild  spacetime. In this case, the condition that the time and space cross terms vanish in the Finsler function no longer holds if the Finsler spacetime is to describe all acceleration scales.

\begin{figure*}
\begin{center}
\includegraphics[width=50mm]{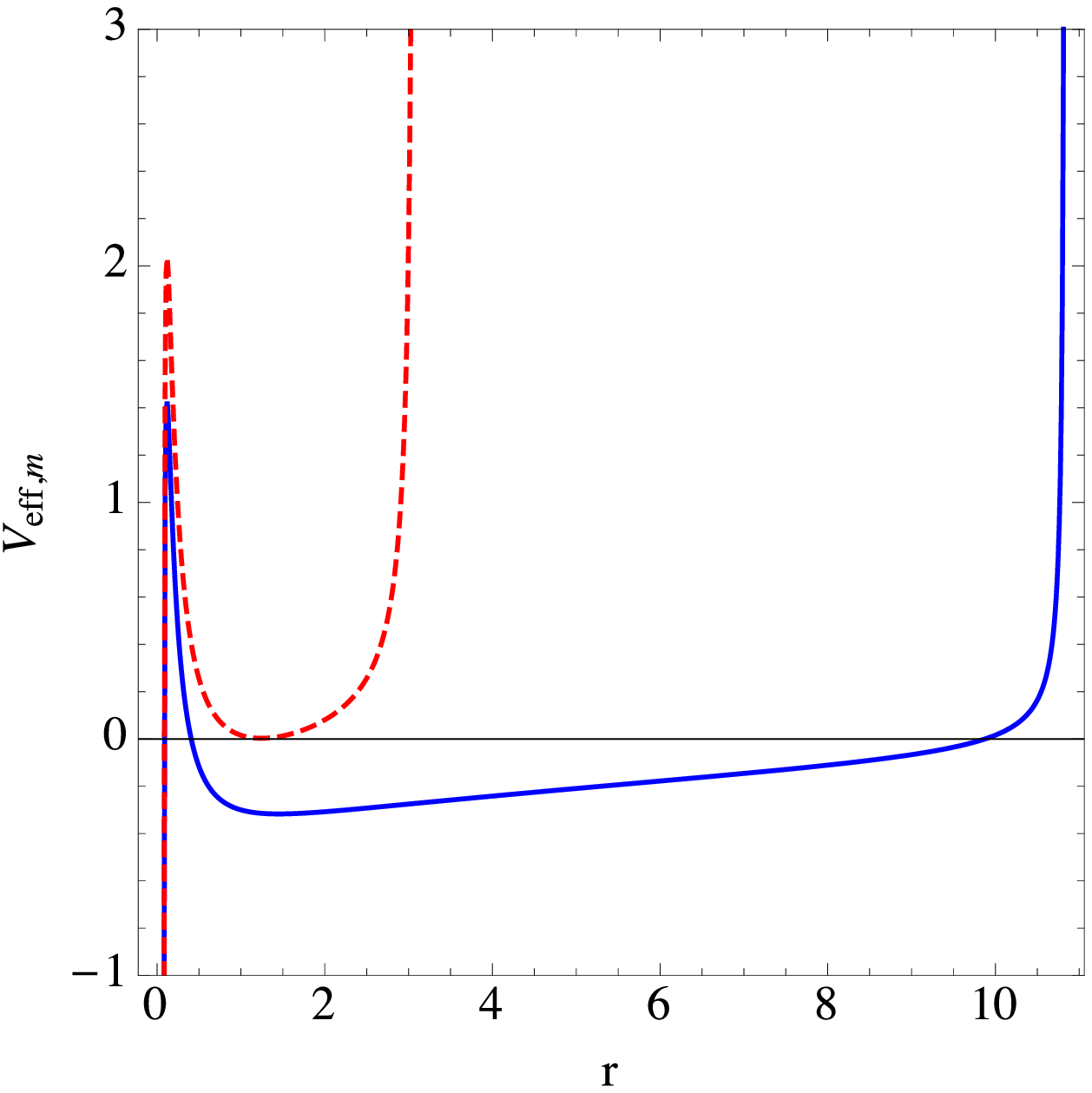}\hspace{10mm}\includegraphics[width=50mm]{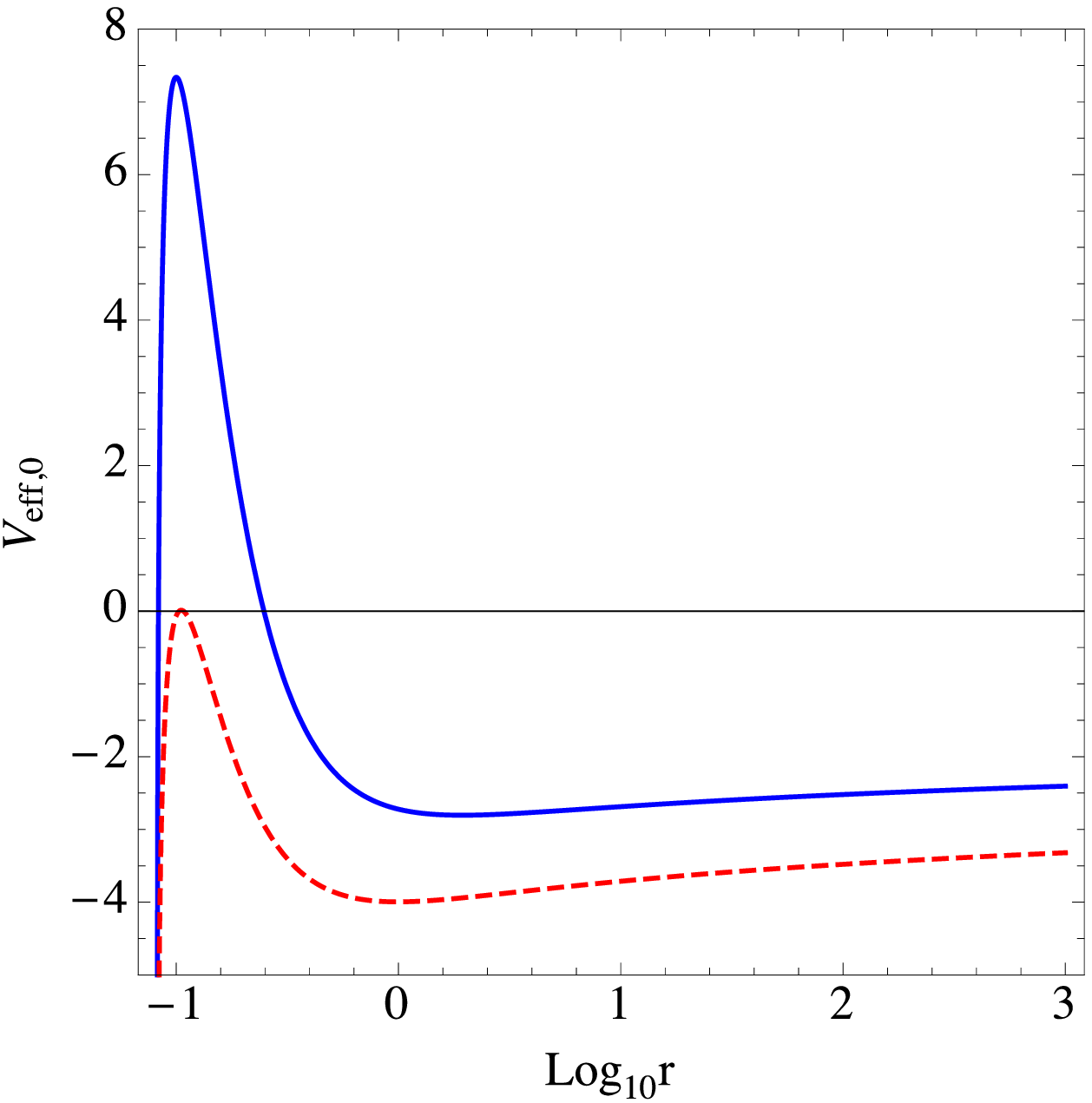}
\caption{Effective timelike (left) and null (right) potentials. For the former, a circular orbit (dashed) and an eccentric orbit (solid) are shown with $(Ma_0,E,J)=(0.1,1.25,0.05)$ and $(0.1,1.132,0.05)$ respectively. For the later, an unstable circular orbit (dashed) and an unbound orbit (solid) are shown with $(Ma_0,E,J)=(0.1,17.15,2.5)$ and $(0.1,9,2.5)$ respectively.}
\end{center}
\end{figure*}

The null effective potential, $V_{\rm eff,0}$, of a point mass is shown in Figure (1,right). There are two equilibria, one stable and another unstable. Motion may be unbound. As with timelike motion, the unstable equilibrium is  reminiscent of that of the Schwarzschild null potential. To ascertain the nature of the stable point we determine the radial motion extrema given by $v_r=0$.  These occur at the distance  $r_m^4=b^4 [1+ 4 \bar M \log (r_m/r_0)] $ where $\bar M=M a_0$ and $b=J/E$ that may be termed the impact parameter.  There are only two solutions that may be written as:
\begin{eqnarray} 
r_m&=& b\bar M^\frac{1}{4} \left[-W_0\left(-\frac{r_0^4e^{-\frac{1}{\bar M}}}{b^4\bar M}\right)\right]^\frac{1}{4} \label{Sol1},\\
r_m&=& b\bar M^\frac{1}{4}\left[-W_{-1}\left(-\frac{r_0^4e^{-\frac{1}{\bar M}}}{b^4\bar M}\right)\right]^\frac{1}{4} \label{Sol2},
\end{eqnarray}
where $W_0$ and $W_{-1}$ are the two branches of the real Lambert's function whose argument satisfies $-e^{-1}\leq -r_0^4b^{-4}e^{-(1/\bar M)}/\bar M\leq 0$ implying $b>r_0  \bar M^{-1/4}e^{(\bar M-1)/4\bar M}$. The two solutions  correspond to  unbound motion, one of which (\ref{Sol1}) is unphysical since the photon is reflected at the larger $r_m$ value (\ref{Sol2}). As the impact parameter increases, the unphysical solution tends to the zero of the time potential $\beta$.  The apparent stable minimum is therefore never reached regardless of energy and angular momentum, implying that motion is unbound. We discuss light deflection in the next section.

We now turn to the spacetime defined by the quartic length function $F_q$ with $\gamma\neq\alpha^2$. The radial velocity is the solution of the polynomial equation of degree 6 in $v_r^2$ : 
\begin{equation} 
\left(1-\frac{(\gamma -\alpha^2)v_r^4}{\beta^{-\frac{1}{3}} E^\frac{4}{3}-\epsilon}\right)\gamma v_r^2= 
\left[\frac{(\alpha^2-\gamma)J^4}{r^4(\beta^{-\frac{1}{3}} E^\frac{4}{3}-\epsilon)^2}  +\frac{ \gamma \left[\beta^{-\frac{1}{3}} E^\frac{4}{3}-\epsilon -(\gamma-\alpha^2) v_r^4\right]^2}{\beta^{-\frac{1}{3}} E^\frac{4}{3}-\epsilon}\right]^\frac{1}{2}
-\frac{\alpha J^2 }{r^2(\beta^{-\frac{1}{3}} E^\frac{4}{3}-\epsilon)}, \label{gen-quartic-pot}
\end{equation}
where $\epsilon=0$ or 1 for null or timelike geodesics respectively, and $v_r=dr/d\tau$ for massive particles. The new effective potential is an implicit function of the gravitational potentials. However, the weak field time potential $\beta$ is still given by equation  (\ref{gravquartic}). To see this explicitly, one may expand the energy constant of massive particles  for both $v_i\ll1$ where $v_i=d x_i/dt$, and for small mass by writing  $\alpha=1+ M \bar\alpha$, $\beta=1+ M \bar\beta$ and  $\gamma=1+ M \bar\gamma$, to find:
\begin{eqnarray} 
E&=&
1+\frac{3(v_r^2+r^2v_\phi^2)^2}{4} + \frac{M\bar \beta}{16}+ \frac{M}{16}\,\left[3(4\bar\gamma-3\bar\beta)v_r^4 +6(4\bar\alpha-3\bar\beta)r^2v_\phi^2v_r^2-9\bar\beta r^4v_\phi^4\right], \label{expquartic}
\end{eqnarray}
and $J=r^2v_\phi (v_r^2+r^2v_\phi^2)$. weak field motion depends only on the time potential $\beta$ through the  third term in equation (\ref{expquartic}) except for the additional perturbation effects such as bound orbit precession given by the last term. Using the Tully-Fisher relation,  the weak field gravitational equation (\ref{gravquartic}) may be recovered easily. The features examined earlier for the quartic metric with $\gamma=\alpha^2$ remain generally valid for massive particles when  $\gamma\neq\alpha^2$.

 \subsection{Rational metrics}
The general static isotropic rational length function may be written as:

\begin{equation}
F_r^2=\left(\frac{{\rm d}\tau}{{\rm d}\lambda}\right)^2=\frac{\beta_0 v_t^4 - \beta_1 v_r^4-\beta_2 r^4 v_\Omega^4 - \beta_3 r^2 v_\Omega^2 v_r^2  + \beta_4 r^2 v_\Omega^2 v_t^2 +  \beta_5 v_r^2 v_t^2}{v_t^2+ \alpha_1 v_r^2 +\alpha_2 r^2 v_\Omega^2  },
\end{equation}
where $v_\mu=dx^\mu/d\lambda$ and the potentials $\alpha_i$  and $\beta_i$  are functions of the radial variable $r$ only. Although unusual, a rational metric offers an interesting possibility with respect to quartic metrics. It may be possible to  interpret the proper length, $d \tau$, as the ratio of quantity associated with a surface area to a  length element.  More specifically, the length function's numerator may be  related indirectly to surface area through the Fresnel 4-tensor of an area-metric spacetime (see Section 5).  The obvious drawback of the rational metric is the larger number of unknown potentials that we reduce by requiring a quartic dependence for energy and cubic for angular momentum.

As with the quartic metric, we set the orbital plane to  $\theta=\pi/2$. The linear dependence of angular momentum $J=-\partial_{v_\phi} F_r$ on $v_\phi$ is eliminated if $\beta_4=\beta_0 \alpha_2$ as can be shown by simple algebra.  Requiring that the quadratic radial velocity term vanishes (per velocity isotropy) in the energy constant  $E=\partial_{v_t} F_r$ leads to $\beta_5=\beta_0 \alpha_1$. The length function then simplifies to:
\begin{equation}
F_r^2=\beta_0 v_t^2 - \frac{ \beta_1 v_r^4 +\alpha_2 r^4 v_\phi^4 + \beta_3 r^2 v_\phi^2 v_r^2 }{v_t^2+ \alpha_1 v_r^2 +\alpha_2 r^2 v_\phi^2 }. \label{Fr}
\end{equation}
where $\beta_2=\alpha_2$ was set by choosing the variable $r$ to mean the physical  spherical radius at constant radius and time as was done  for the quartic metric. We note that the relations among  $\alpha_1$, $\alpha_2$, $\beta_0$,  $\beta_4$ and $\beta_5$ no longer hold if the small acceleration domain is left and the Finsler function is to describe all acceleration scales and tend to a  Schwarzschild metric.

The Tully-Fisher relation may be implemented by considering the weak field limit, $v_{\rm circ}\ll 1$, for which the constants of motion may be expanded to fourth order in velocity to yield the geodesic equations as follows:
\begin{eqnarray}
E&=&\beta_0^{1/2}+ \frac{3(\beta_1v_r^4+ \beta_3 r^2 v_r^2 v_\phi^2+\alpha_2r^4 v_\phi^4)}{2\beta_0^{1/2}}, \\
J&=&    \frac{ r^2 v_\phi( \beta_3 v_r^2 +2 \alpha_2 r^2 v_\phi^2)}{\beta_0^{1/2}},   
\end{eqnarray}
where $v_i=dx^i/dt$. Note that the  potential $\alpha_1$ is absent from the equations of motion. The solution for the radial velocity $v_r$ is unsurprisingly similar to that of the general quartic metric  (\ref{gen-quartic-pot}):
\begin{eqnarray}
2\beta_1\left[ 1 -\frac{3\left(\beta_1-{\beta_3^2}/{4\alpha_2}\right)v_r^2}{2\left(\beta_0^\frac{1}{2}E-\beta_0\right) }\right] v_r^2&=&
\left[\frac{9(\beta_3^2-4\beta_1\alpha_2)\beta_0^2J^4}{4\alpha_2^2r^4\left(\beta_0^\frac{1}{2}E-\beta_0\right)^2}+  \frac{8\beta_1}{3}\, \left(\beta_0^\frac{1}{2}E-\beta_0\right)\left(  1 -\frac{3\left(\beta_1-{\beta_3^2}/{4\alpha_2}\right)v_r^2}{2\left(\beta_0^\frac{1}{2}E-\beta_0\right) }\right)^2\right]^\frac{1}{2} \nonumber \\ && -\frac{3J^2\beta_0\beta_3}{2\alpha_2r^2\left(\beta_0^\frac{1}{2}E-\beta_0\right)}.
\end{eqnarray}
In the case where $\beta_3^2=4\beta_1\alpha_2$ (the equivalent of $\gamma=\alpha^2$ for the quartic metric), the effective potential simplifies to: 
\begin{eqnarray}
&&2\beta_1^\frac{1}{2} v_r^2=\left(\frac{8}{3}\right)^\frac{1}{2} \left(\beta_0^\frac{1}{2}E-\beta_0\right)^\frac{1}{2}-\frac{3J^2\beta_0}{\alpha_2^\frac{1}{2}r^2\left(\beta_0^\frac{1}{2}E-\beta_0\right)}\equiv -V_{\rm eff,m}\ ,
\end{eqnarray}
and a relation between the potentials $\beta_0$ and $\alpha_2$  for $v_{\rm circ}\ll 1$ may be determined  by examining the circular orbits of $V_{\rm eff,m}$ as it was done for the quartic potential. If it is further assumed that $\alpha_2$ is constant, the time potential $\beta_0$ satisfies a differential equation similar to that of the quartic potential $\beta$ (\ref{gravquartic}) showing that $\beta_0\propto  M a_0 \log (r/r_0)$ for an isolated mass.

\section{Light deflection}
 Our program of searching for static isotropic Finsler spacetimes for MOND based on the Tully-Fisher relation has so far led only to the determination of the metric's time potential in the weak field approximation. Space potentials may be constrained from light deflection as the process is directly influenced by the space curvature of the metric (see below). For instance, in the Schwarzschild metric of an isolated mass in a Riemann spacetime, light deflection in the weak field limit is due in equal measure to time and space curvatures.  In what follows, we examine light delfection of MOND's Finsler metrics for small accelerations in the weak field approximation.  The small acceleration approximation does not mean that photons are not relativistic. Similarly, the weak field approximation  means only   that the perturbation that deflects the free relativistic motion of photons is small \citep{Schneider92}.

For the quartic metric, light travels on the null surface given by $F_q=0= \beta v_t^4 - (\gamma-\alpha^2) v_r^4 - ( \alpha v_r^2 + r^2 v_\phi^2)^2$ at constant energy 
$E= \beta v_t^3$ and constant angular momentum $J=r^2v_\phi( \alpha v_r^2 + r^2 v_\phi^2) $. Its orbit $\phi(r) $ may be obtained from the three relations by eliminating $v_\phi$ as:
 \begin{equation} 
{J^4\beta}{(\gamma  \phi^{\prime -4} + r^4  + 2\alpha  r^2  \phi^{\prime -2})^3}={E^4 r^8}{ ( \alpha \phi^{\prime -2} + r^2 )^4},\end{equation}
where the prime denotes $\partial_r$. If the light surface is quadratic ($\gamma=\alpha^2$) then the deflection equation is identical to that of Riemann spacetimes.  Otherwise ($\gamma\neq\alpha^2$), $\phi^{\prime -2}$ is the solution of a polynomial of degree 6. For impact parameters  $b=J/E>r_0e^{-1/4Ma_0}$, we may use the closest approach distance, $r_m$ defined as  $dr/d\phi(r_m)=0$ and the notation  $\beta_m=\beta(r_m)$ to cast the deflection equation as:
 \begin{equation} 
{\beta r_m^4}{(\gamma  \phi^{\prime -4} + r^4  + 2\alpha  r^2  \phi^{\prime -2})^3}={\beta_m r^8}{ ( \alpha \phi^{\prime -2} + r^2 )^4}. \label{defl1}
\end{equation}
We determine light deflection with the assumption that the potentials $\alpha$ and $\gamma$ remain constant as mass vanishes and spacetime may be described by a flat Lorentz metric in that limit. Writing
$\alpha=1+ \bar \alpha $,  $\beta=1+ \bar \beta $,  $\gamma=1+ \bar \gamma $, $\beta_m=1+ \bar \beta_m $ where $\bar \alpha,\bar \beta, \bar \gamma\ll 1$, and expanding (\ref{defl1}) to first order with respect to the perturbation, the deflection equation reads:
\begin{eqnarray} 
 \phi^{\prime} &=& \frac{1}{r_m u(u^2-1)^{1/2}} + \frac{(  \bar\beta  - \bar\beta_m) u  }{4r_m (u^2-1)^{3/2}}+
  \frac{  (3\bar\gamma - 4 \bar\alpha) u^2+ 3(2\bar\alpha-\bar\gamma) }{4r_m u^3(u^2-1)^{1/2}}, \label{defl2}
\end{eqnarray}
where $u=r/r_m$. The three terms correspond respectively to rectilinear motion in flat empty  space,  deflection caused by  time curvature and  deflection caused by space curvature. 

Applied to the quartic Finsler spacetime of an isolated mass, the deflection produced by the time potential $\bar\beta =4 a_0 M \log (r/r_0)$ is $\Delta \phi=\pi Ma_0/2$ similar to that from a singular isothermal sphere: linear with respect to mass and  independent of the characteristic scale $r_0$. Time curvature of the Finsler metric therefore cannot account for the larger observed deflections. For comparison, interpretations of MOND based on gravity  modification provide a  deflection scaling $\Delta \phi\propto (Ma_0)^{1/2}$  \citep{Milgrom13,Mortlock01} that explains some but not all of the observed mass discrepancy in galaxy clusters. For Finsler metrics,  space curvature may account for a larger deflection in particular because of its possible dependence on the distance scale $r_0$ that in turn may depend on $M$ and possibly $a_0$. For instance, if  we  arbitrarily set $\bar \alpha = c_\alpha M a_0\log(r/r_0)$ and  $\bar \gamma = c_\gamma Ma_0\log(r/r_0)$, deflection is given by:
\begin{equation}
\Delta\phi= \frac{\pi M a_0}{8} \left[4+c_\gamma-2c_\alpha + (2 c_\gamma - 4 c_\alpha/3)\log(b/2r_0)\right]. \label{arbitdef}
\end{equation}
In terms of the two distance scales of the system $r_u=a_0^{-1}$, a constant related to the radius of the observable Universe, and $r_s = 2 M$, the Schwarzschild radius, deflection  $\Delta\phi\propto r_s\log(b/2r_0)r_u^{-1}$ for large impact parameters. The proportionality constant must be positive if deflection is to be convergent; this in turn, depends on the definition of $r_0$ whether it is larger or smaller than the intermediate acceleration radius $r_e=(r_ur_s/2)^{1/2}$ where the Newtonian gravitational acceleration is comparable to $a_0$.  Substituting galaxy-galaxy lensing typical parameters show that the arbitrary form of $\bar\alpha$ and $\bar\gamma$ that yields (\ref{arbitdef}) does not produce sufficient lensing whether $r_0$ is taken of the order of $r_s$ or $r_u$. Obviously, this  indicates only that the arbitrary choice of the space potentials is inadequate. Light deflection may constrain but not determine the space potentials as the effect essentially probes the integral potentials and not the corresponding acceleration through such integral equations as (\ref{defl2}).  We note that the deflection equation for the rational metric (\ref{Fr}) is quite involved but its linearization is similar to (\ref{defl2}) and the time potential $\beta_0$ yields a constant deflection for an isolated mass under the assumptions specified at the end of Section 3.2. The source of larger deflection for the rational metrics is also space curvature.
 
\section{Discussion}
In our search for an interpretation of MOND as a modification of inertia at small accelerations, our starting point was Milgrom's original modification of Newton's second law for circular motion. We have been able to show that implementing the Tully-Fisher relation in that case is simple but comes at a price. Although the gravitational potential remains linear with respect to mass, it does not satisfy the Poisson equation in the weak field limit as one would think from Milgrom's original scaling by equating $v^4r^{-2}a_0^{-1}$ and  $Mr^{-2}$. The  new gravitational equation (\ref{gravquartic}) differs from a Poisson equation in that the divergence is not applied to the gradient of the potential $\beta$ but to the product of $\partial_r\beta$ and $r_ur^{-1}$ where $r_u=a_0^{-1}$ is proportional to the radius of the observable Universe. There is also a difference in the meaning of the Finsler and Riemann  gravitational time potentials in that the former  is not equivalent to the latter but to the latter's square as may be seen on the quartic metric with a quadratic light surface (\ref{quarticmetric2}). Although unexpected, the necessary modification of gravity preserves the linear character of the gravitational potential which may prove a precious advantage with respect to modified gravity theories. Moreover, covariant theories of modified gravity MOND require various auxiliary fields in addition  to the physical metric in order to implement the Tully-Fisher relation. In this sense, our modification of inertia and its corresponding linear  modification of gravity are minimal.

Another element that appears in our analysis of asymptotic motion is the characteristic distance $r_0$. Whereas it does not influence weak field massive motion, it is likely to contribute to light deflection through its possible dependence on mass and  the acceleration scale $a_0$. A covariant MOND formulation possesses two independent distance scales: $r_u$, a multiple of the radius of the observable Universe, and  $r_s$, the Schwarzschild radius. It is not clear yet how the distance $r_0$ depends on  these scales but its form should be related either to the transition from the small to the large acceleration domains or perhaps, more interestingly, to the presence of a coupling of local dynamics to large scale structure evolution through the acceleration scale $a_0$. In this regard, our approximation of static and asymptotically  flat metrics (section 4)  that implements the Tully-Fisher relation locally for weak field massive motion may not be adequate for light propagation. Although complicated by the absence of a full modified-inertia gravitational equation, it would be desirable to determine Finsler metrics for MOND equivalent to  the McVittie-type solutions of the Einstein gravitational equation for a mass embedded in an expanding Universe \citep{McVittie33,Nandra12}.

The appearance of the distance scale $r_0$ has another implication  regarding the possible scale invariance from which MOND may emanate in the domain of small accelerations.  \cite{Milgrom09,Milgrom14} showed that the basic phenomenological MOND laws are natural consequences of the assumption  that the equations of motion of purely gravitational systems are invariant under the change of $(t,{\bf r})$ into $(\lambda t,\lambda {\bf r})$ when the acceleration scale $a_0$ tends to infinity.  
Our formulation that is based on implementing the Tully-Fisher relation indicates that the converse may not be true. 
In effect, the time potential we derived  $\beta(r)=1+4Ma_0 \log(r/r_0)$ (for the quartic metric) enters the geodesic equations directly through the Finsler function so that the equations of motion are not scale invariant. The distance $r_0$ appears in various places: it is related to the distance at which the metric becomes singular ($\beta=0$), although that distance is located far inside the Newtonian domain where the Finsler metric in its MOND form does not apply (lacking space and time cross terms).  The distance $r_0$ also enters the expressions of the energy, angular momentum, and (proper) period. Light deflection may also depend on $r_0$ and in this respect our formulation departs significantly from modified-gravity MOND theories where asymptotic light deflection is constant akin to that of dark matter isothermal sphere models albeit with a different mass dependence. 
 Constant asymptotic light deflection is sometimes considered a tenet of MOND phenomenology but  its observational validity is questionable when applied to galaxy cluster encounters. We surmise that the appearance of the distance scale $r_0$ is ubiquitous in the context of MOND as whatever type of interpretation is considered (gravity modified or inertia modified) when formulated in a metric setting is bound to have an asymptotic metric that requires $r_0$ because of the logarithmic behavior of the gravitational potentials.  Once the scale $ r_0$ enters a metric, the geodesic equations of motion obtained from the proper length action may no longer be scale invariant and light deflection will depend on the characteristic scale $r_0$.\footnote{The Riemannian metric of an isothermal sphere is logarithmic but the corresponding light deflection is constant. This peculiar property is related  to the degeneracy of the metric in the weak field approximation.  The  metric is commonly  written in that approximation using isotropic coordinates (in gravitational lensing studies in order to apply the Fermat theorem) but when written in the more physical Schwarzschild-like coordinates where the radial coordinate denotes the physical radius at constant time and radius (similar to the coordinates we use in this paper), one finds that, consistant with the  weak field approximation, the space potential is constant and independent of $r_0$. Since the time potential contributes to deflection through its first derivative and the space potential contributes through its expression  (underived), the scale $r_0$ is absent. Excepting such particular potentials, asymptotically logarithmic potentials should yield  light deflections that depend on the distance scale $r_0$.}

The simplest Finsler metrics that are compatible with the Tully-Fisher relation make use of polynomials of degree 4 with respect to local displacements. The classical physical setting where such expressions appear is surface area measurements albeit it as the norms of  differential 2-forms. As the corresponding tensors are antisymmetric, we may look for symmetric 4-tensors in the context of area-metric spacetimes. An area metric is an invertible partly antisymmetric covariant tensor field, $G_{\kappa\lambda\mu\nu}$, satisfying: $G_{\kappa\lambda\mu\nu}=G_{\mu\nu\kappa\lambda}$, $G_{\kappa\lambda\mu\nu}=-G_{\lambda\kappa\mu\nu}$ and $G^{\kappa\lambda ab}G_{ab\mu\nu}=2(\delta^\kappa_\mu\delta^\lambda_\nu-\delta^\kappa_\nu\delta^\lambda_\mu)$. It  gives rise to the totally symmetric Fresnel tensor field ${\cal G}_{\mu\nu\kappa\lambda}$ defined by 
\begin{equation}
{\cal G}_{\mu\nu\kappa\lambda}= -\frac{|{\rm det}\bar G|^{-\frac{1}{3}}}{24}\epsilon^{ijkl}\epsilon^{mnpq} {G}_{ijm(\mu}G_{\nu|kn|\kappa}G_{\lambda)lpq},
\end{equation}
where  $\epsilon_{\mu\nu\kappa\lambda}$ is the totally antisymmetric tensor and $\bar G_{\mu\nu\kappa\lambda}=
G_{\mu\nu\kappa\lambda}-G_{[\mu\nu\kappa\lambda]}$ is the cyclic part of the area-metric tensor. In a Riemann spacetime with a metric tensor $g_{\mu\nu}$, the naturally induced area-metric tensor $G_{\mu\nu\kappa\lambda}=g_{\mu\kappa}g_{\nu\lambda}-g_{\mu\lambda}g_{\nu\kappa}$, and the Fresnel tensor simplifies  to ${\cal G}_{\mu\nu\kappa\lambda}=g_{(\mu\nu}g_{\kappa\lambda)}$ \citep{Punzi07}.  The Fresnel tensor first appeared in the study of electromagnetism in anisotropic media \citep{Tamm24,Tamm25,Rubilar02} and  particularly of birefringence \citep{Obukov02} as well as in the study of area-metric spacetimes that do not possess a prior metric tensor \citep{Punzi09b}. Light propagation in the geometrical optics limit is shown to occur (in anisotropic media and area-metric spacetimes) on the null quartic surface   defined by ${\cal G}^{\mu\nu\kappa\lambda} k_\mu k_\nu k_\kappa k_\lambda$  where $k_\mu$ is the wave covector \citep{Hehl}. The Fresnel tensor in an area-metric spacetime is therefore a possible candidate for constructing a Finsler distance function for MOND. The study of massive motion in area-metric spacetimes is trickier than light propagation as the fundamental objects on such spacetimes are not particles but strings \citep{Schuller06}. \cite{Punzi09a} devised an ad hoc method for defining a localized massive particle by averaging the motion  a string fluid and showed that such motion is described by the geodesics of a quartic metric defined by the Fresnel tensor. Understanding massive motion in area-metric spacetime is a first step towards determining the physical conditions that give rise to MOND-like inertia and how it is related to large scale structure evolution.

\section*{Acknowledgments}
The author thanks Mordechai Milgrom for useful comments on the manuscript.                                                                                                                                                                                                                                                                                                                                                                                                                                                                                                                                                                                                                                                                                                                                                                                                    

\bibliographystyle{mn2e}

\bibliography{sms}

\end{document}